\documentclass[12pt]{article}
\usepackage[dvips]{graphicx}
\usepackage[dvips]{epsfig}

\date{}

\begin{document}

\begin{center}

\textbf{Modulation of Autonomic Nervous System activity by Gyrosonic
stimulation
}

\vspace{1cm}

\textbf{S.K.Ghatak$^{a,*}$, B.Roy$^{b}$, R.Choudhuri$^{c}$ and S.Bandopadhaya $^{c}$} \\
$^{a}$Department of Physics and Meteorology, $^{b}$School of Medical Science and Technology\\Indian Institute of Technology,Kharagpur-721392,India\\
$^{c}$ WINGARD, Institute of Visual and Auditory Research, Kolkata 700012 ,India\\

\end{center}

\vspace{0.5cm}

\begin{abstract}

Gyrosonic is a novel audio binaural stimulus, that generates rotational perceptions of sound movement in the brain at a particular predetermined frequency. The influence of gyrosonic on the  autonomic nervous system of healthy subjects has been examined by analyzing heart rate variability (HRV) in time- and frequency- domain. The M-lagged Poincare plot shows that the parameters SD1, SD2 and ratio SD12 (SD1/SD2) increases with lagged number M, and M-dependence is well described by Pade$\acute{}$ approximant $\chi \frac{1+\beta M}{1+\gamma M}$  where values of $\chi$, $\beta$ and $ \gamma$ depends on parameters SD1, SD2 and SD12. The values of parameters SD1, SD2 and SD12 after gyrosonic stimulation are augmented, compared to pre-stimulation values for all M. The slope and curvature that define the variation of SD12 with M increase considerably due to the stimulation.Slowing down of the Heart rate and increase in the standard deviation SD and the root-mean squared successive differences (RMSSD) in the RR-interval are observed after stimulation. The spectral power of both low (LF) - and high (HF) - frequency go up due to the stimulation.The DFA analysis of RR interval exhibits a decrease in value of exponent $\alpha$ due to the gyrosonic. The results indicate that there may be an improvement of the sympatho-vagal balance due to this novel audio stimulus.

\end {abstract}

\vspace{1.0cm}


\vspace{0.5cm}

%\noindent\textbf{PACS}:71.10.Fd, 75.30.Mb

\vspace{2.0cm}

\noindent $^*$Corresponding author.

\noindent E-mail address: skghatak@phy.iitkgp.ernet.in

\newpage

\newcommand{\be}{\begin{equation}}
\newcommand{\ee}{\end{equation}}

\section{Introduction}
\label{Introduction}

Autonomic nervous system (ANS), with its main two divisions; Sympathetic and Parasympathetic, may be viewed as a hierarchically coordinated neuronal network. It regulates the exchange of energy and information with the environment [1]. Most of the internal organs like the heart, gastrointestinal tract, lungs, urinary bladder and blood vessels are under the influence of the autonomic nervous system. Usually the sympathetic and parasympathetic nervous systems have opposing effects on the organs. The degree of influence may vary and this determines the ultimate activity of that organ. The sympathetic nervous system is rapidly activated in physical or mental stressful conditions. It increases the heart rate, cardiac output and blood flow to the muscles. It dilates the pupil and decreases gastrointestinal tract activity. Its action is therefore, sometimes aptly referred to as the 'flight or fight' response. On the other hand, the parasympathetic system decreases the heart rate and blood pressure, constricts the pupil and increases gastrointestinal activity. The 'rest and digest' response is coined for its action[2]. The body maintains a proper balance between the sympathetic and parasympathetic activity for appropriate functioning. A deviation from this balance may result in disease conditions like acute coronary syndrome, chronic heart failure, diabetes mellitus etc [3].\\
It has been shown that sensory inputs, like a harmonic auditory stimulus, can have a wide range of psychological and physical effects [4]. A sensory input can trigger a cascading series of events along the nerve down which it travels [5]. Many of the beneficial effects of the auditory stimulus take their origin along the route of the impulse. Music has the capacity to modify the psychobiological state of human. It can thus relieve stress and stress related ailments [6].Gyrosonic is a novel auditory stimulus and appears to possess such capacity. When applied binaurally, it produces the perception of rotation of the audio source inside the brain. The frequency of this rotation is in the infrasonic region ($\sim2$Hz). The perception of movement in the auditory space by human depends on a number of cues.It has been demonstrated that the moving sound produced by the sequential excitation of a mono source through several speakers in a free field, can generate a specific activity in the brain [7]. Sound motion has been shown to evoke a magnetic field inside the brain,and the evoked magnetic response is specific to moving sound and is absent for non-moving sound stimulus [8]. Another study has shown that the right parietal cortex is involved in the processing of sound motion [9]. Compared to the earlier experiments gyrosonic stimulus has some improved rotational features . Thus it can be expected that this may produce a larger spectrum of brain activation. Earlier studies with this moving sound showed that the arousal level of psychosomatic patients was significantly reduced [10].

The heart rate variability (HRV), which is a measure of the beat to beat fluctuation of the heart rate, reflects the time varying influence of ANS and its components on the cardiovascular function. Due to its non-invasive nature and convenience of the measurement, the HRV is often used to assess the influence of the autonomic nervous system on the cardiac function. Under normal circumstances the HRV is regulated by the combined action of heart automaticity and the feedback elements from the vagal and sympathetic activity on the heart. The stimuli that are capable of altering the feedback components therefore change the dynamics of the heart. Gyrosonic stimulus is assumed to be one such stimulus. In this paper we have estimated the influence of gyrosonic stimulus on the dynamics of the heart by assessing the Heart Rate variability (HRV) in response to this stimulus.The HRV can be assessed in time domain and in frequency domain. The time domain analysis provides quantitative measurement of the variation of heart rate, standard deviation SD, root mean square successive differences (RMSSD) in RR-interval.On the other hand the frequency
domain analysis of HRV mainly measures the high frequency (HF) and low frequency (LF) power spectrum of fluctuation of the RR interval. It has been suggested that the high frequency power spectrum is modulated by parasympathetic (mainly vagal) activity and the low frequency power spectrum is mainly influenced by both parts of ANS activity [11,12].

Dynamics of the cardiac system is nonlinear. Hence it is reasonable to assume that a nonlinear analysis is more appropriate means to get an accurate idea about the cardiac system. Poincare$\acute{}$ plot analysis meets this criterion [13, 14, 15].  It is a nonlinear geometric method of heart rate analysis. It is basically a scatter plot of any heart rate interval $RR_n$ and next one $RR_{N+1}$.When this plot is adjusted with an ellipse,three important parameters then define the plot.These are the length of the semi-minor axis of the ellipse SD1, the length of the semi-major axis of the ellipse SD2, and their ratio SD12.The parameter SD1 is the standard deviation of instantaneous beat to beat heart rate variability and it is the measure of short term variability of heart. It is mainly influenced by parasympathetic regulation on the heart. Other parameter SD2 is the measure of long term variability. It has been shown that these parameters are correlated with the power spectral density of heart rate fluctuation. SD1 is correlated with the high frequency spectrum and SD2 is mainly  associated with the low frequency spectrum [16].Instead of taking the successive RR interval, the Poincare$\acute{}$ plot can be generalized by taking a lag of greater than $1$ between RR intervals. A study with a lag of $4$ has shown evidence of being superior to normal Poincare$\acute{}$ [17]. Moreover, a study on chronic heart failure patients has demonstrated strong dependency of the parameter SD12 on the lag interval. This study has also shown that the curvature of the plot of SD12 with various lag numbers is significantly different in patients and normal subjects. In chronic heart failure patients the curvature is much smaller [18]. Another study on diabetic patients has pointed out that the SD1 decreases in diabetic patients with higher lag numbers [19]. Using this lagged Poincare$\acute{}$ plot it has been reported that SD1 and SD2 decreases with an increase in the lag number for smokers [20].
Apart from the spectral analysis and Poincare$\acute{}$ plot analysis for the heart rate variability, another method, called Detrended Fluctuation Analysis, can be applied with some added advantages. It has been found that under healthy conditions, the RR interval time series exhibits long-range power law correlation. Similar behavior has also been found in other physical systems [21]. The Detrended Fluctuation analysis proposed by Peng et al is supposed to reflect on this long range correlation of the RR interval time series [22, 23]and is widely used.

In our study we have used all these methods to examine the change in HRV in healthy subjects after the gyrosonic stimulation. Through this study we propose to show that a change in the autonomic regulation occurs due to the influence of the gyrosonic stimuli.

\section{Stimuli}

In this study we have used a gyrosonic stimuli constructed from an Indian percussion instrument, the Tabla. The stimuli were recorded digitally at a sampling rate of $44.1$ kHz and in $16$ bit. Then the amplitude of sound was modulated at $2$ Hz in such a way that the amplitude increased in one ear and decreased in the other ear. This modulation created a perception of moving sound. When one ear heard the sound with higher amplitude it assumed the sound to be moving towards that ear whereas the ear hearing the lower amplitude sound took that the sound was moving away. This produced the effect as though the sound was moving in a horizontal plane around the head. There was a phase of $0.568$ seconds between the advancing or the rising and receding or the falling stimuli. The slopes of the amplitude of the rising or falling sound were exponential. The stimulus was presented through headphones, connected to a computer through a custom electronic interface. The sound was played at a sensation level around $50$ db. The ultimate perception produced was that of a rotating sound moving in a horizontal plane inside the brain. The duration of the stimuli was $9.5$ minutes. This duration was chosen to ensure that the subject perceived the effect in its optimum potential. No subject experienced nausea or episode of vertigo during or after the playback of the stimuli. The stimuli were given in a sound proof room with the subjects in recumbent position.

\section{Subjects and Measurements}

Thirty one subjects ($11$ male, $20$ female with average age $36\pm 12$ yrs)  volunteered for the study. All of the subjects were explained about the study and all of them gave their consent for the study. All the subjects were healthy and were not on any kind of medication. There was no history of any type of nervous system disorder. The study was approved by the ethical committee of IIT Kharagpur, India.
ECG data was taken from three limb leads. Patients were in supine condition. Sampling rate was $500$ Hz with a resolution of $12$ bit. A pre stimuli ECG was taken for $10$ mins. Then the subjects listened to the gyrosonic stimulus. The post stimulation ECG for $10$ mins was recorded after a gap of $10$ mins from the end of the stimulation.
Whole $10$ mins data of ECG for both pre and post stimuli was analyzed after selecting the sinus beats only. Ectopic beats were detected visually and deleted manually. RR interval or better to say normal beat to normal beat interval was detected through the Origin software. The non linear analysis like the Poincare$\acute{}$ plot and the Dretended Fluctuation analysis were done with program written in Matlab Software. The linear HRV analysis was done using standard program.

\section{Data analysis}

During the analysis of HRV we have strictly followed the Task Force guideline of HRV analysis[13]. In linear time domain the heart rate (HR), standard deviation (SD) of RR interval (basically the normal beat to normal beat interval), root-mean squared successive differences (RMSDD) were calculated. The ratio of SD/RMSDD was considered to be a good measure of the sympatho-vagal balance [14].
In the frequency domain, the power spectral density for the low frequency ($0.04$-$0.15$ Hz) and the power spectral density for the high frequency ($0.015$-$0.5$ Hz) were calculated. The distribution of power and the central frequency of LF and HF were not fixed. They usually vary in relation to changes in the autonomic modulation of the heart. According to the Task Force Guideline the RMSDD represented the HF variability and the total variability of the RR interval data was reflected by the SD. The parameters SD1 and SD2, as described earlier, were the width and the length of the ellipse. The ellipse was fitted to the Poincare plot of RRN+M  vs RRN where M is the lag number. SD1 and SD2 were calculated for lag M  from the relation :SD1 = $(\Phi(M) - \Phi(0))^{1/2}$ and SD2 = $(\Phi(M) + \Phi(0))^{1/2}$ where  the  auto-covariance  function  $\Phi$(M) is  given by \\
\begin{center}
 $\Phi(M)$ =$E[(RR_N - \bar{RR}) (RR_{N+M} - \bar{RR})]$.
\end{center}
[16].

The long term correlation in RR-time sequence was assessed by Detrended Fluctuation Analysis [22]. The measure of correlation was given by a scaling exponent ($\alpha$) of the fluctuation function $F(\tau)\approx \tau^\alpha$. The computation of fluctuation function $F(\tau)$ was done in the following way. For a given time sequence $R(t_i)$, $t_i = i \delta t$ where $\delta t$ is characteristic time interval for the sequence and  i=1,N , an integrated time series $r(t_i)$ was defined as $r(t_i)=\sum_j^i [R(t_j)-R_m]$, $i=1, N$ where $R_m$ was the mean of $R(t_i)$. The integrated series was divided into boxes of equal size of time  $\tau=n\delta t$ and linear function was used to fit box data. The fluctuation function $F(\tau)$ was calculated as root mean square fluctuations relative to the linear trend. The power law behaviour of $F(\tau)$ provided the scaling exponent. It has been observed that acceptable estimate of the scaling exponent $\alpha$ (from DFA) can be obtained from analysis of data sets of length $256$ samples or greater (equivalent to approximately $3.5$ min for RR data at a heart rate of $70$ bpm). The analysis of RR data for period of $10$ min time interval was therefore expected to provide an adequate measure of the scaling exponent.

\section{Results and discussions}

The results of linear analysis of HRV is summarized in Figs.1. The Fig.1 depicts the
mean value of HR , the mean of SD and RMSSD in RR-interval of all subjects. In all figures presented here the term PRE and PST indicate respectively before and after gyrosonic stimulation. The heart rate was changed from $83.1$  to $81.8$ per min. The slowing down of HR was found to be more for subject with higher initial HR. Both SD and RMSSD were increased in post stimulation. The LF and HF components of spectral power in RR-interval were also increased after stimulation (Fig.1). The increase in LF- power was more than that of HF component. However, the ratio of LF/HF power was decreased in post condition.The changes of all the above parameters are with $p<0.05$

In Fig.2 the lagged Poincare$\acute{}$ plot of a subject was presented with lag of 1,5,9 and 13 plot. The left and right part of the figure represent respectively the situation before and after gyrosonic stimulus. As lag number increased the plot became more scattered with consequent increase in both width and length of the plot. After stimulation the $RR_{N+M}$ vs $RR_N$ plot were more scattered, and center of the plot was shifted to higher value indicating slower heart rate.

The  group average values of parameters SD1, SD2 (both in sec.) and SD12 (SD1/SD2) obtained from corresponding values of individual subject were plotted against lagged number M (Fig.3  points), Both parameters SD1 and SD2 were increasing function of lagged number.After gyrosonic stimulation (PST) their values were higher than those before stimulation (PRE) and the growth rate of SD1 with M is also higher. The result points out that the gyrosonic stimulation can enhance both short and long time correlation of heart beat. The ratio SD12 (points) in post stimulation state was higher than that in pre-state and the difference increased with lag number. The difference between the values of all three parameters before and after stimulation was found to be significant ($p<0.001$). In order to find the relationship of these parameters with lag number M the method of Pade$\acute{}$ approximant [23] was used. Assuming simple form of the Pade$\acute{}$ approximant for SD's as\\

\begin{equation}
Y=\frac{a+bM}{c+dM}= \chi \frac{1+\beta M}{1+\gamma M}
\end{equation}
the ratio of polynomial in M of degree one.  Here Y = SD1, SD2 or SD12 and $\chi = a/c$, $\beta = b/a$ and $\gamma = d/c$ were taken as the new unknown parameters. The above equation was chosen by examining trend for small M and large M. For small M, Y increased linearly and deviated  at higher M. As shown below, the equ.(1) was found to be an excellent representation of the observed dependence of SD's on M (solid line in Fig.3).  When expressed for small M the equ.(1) can be approximated as $Y = C + L M + Q M^2$ where L =$\chi (\beta - \gamma)$ and Q = - $\gamma L$. It is to be noted that such variation of these parameters for small M were also found earlier [18]. The values for L and Q were given in table 1. It was evident that the magnitude of slope and curvature of SD1 after gyrosonic stimulation were increased considerably. On the other hand, the linear coefficient for SD2 increased slightly and curvature remained essentially unaltered. The ratio SD12 for different M for each subject was calculated and the mean value of the ratio was plotted with M in Fig.3 (points in lower curve). The data were excellently fitted by the equation (1) (solid curve) with the parameters value noted in table.1. The ratio SD12 exhibited a larger change in both slope and curvature. The gyrosonic stimulation resulted respectively ${89\%}$  and $44$ percent increase in the magnitude of  curvature  and slope of SD12.

\begin{table}[htb]

\caption{ The value of parameters $\chi$, $\beta$, $\gamma$ obtained from fit of eq.(1) with respective value of $R^2$.The parameters L and Q are the coefficient of linear and quadratic terms in expansion of Y in terms of M. Values of $\chi$, L and Q for SD1 and SD2 are in second. }\

\begin{tabular}{|c|c|c|c|c|c|c|c|}

 \hline
   & a & $\chi$x$10^{-2}$& $\beta$x$10^{-2} $& $\gamma$x$10^{-2} $ & $R^2$ x$10^{-2}$& L x$10^{-3}$& Q x$10^{-4}$\\\hline
  SD1 & PRE & $1.34\pm0.02$ & $25.52\pm0.85$ & $2.15\pm0.11$ & $99.99$ & $3.14\pm0.13$  & $-0.67\pm0.02$  \\\hline
   & PST & $1.68\pm0.02$& $27.25\pm0.96$ & $2.56\pm0.12$ & $99.98$& $4.14\pm0.2$  & $-1.06\pm0.12$ \\\hline
 SD2 & PRE & $3.54\pm0.03$  & $15.07\pm0.56$ & $3.12\pm0.17$ & $99.98$ & $4.23\pm0.31$ & $-1.32\pm0.1$  \\\hline
   & PST & $4.11\pm0.04$ & $13.74\pm0.53$ & $2.94\pm0.17$ & $99.97$ & $4.44\pm0.34$  & $-1.31\pm0.11$  \\\hline
  SD12 & PRE & $38.4\pm0.24$ & $18.43\pm0.73$ & $8.74\pm0.32$ & $99.97$ & $37.2\pm2.2$ & $-32.5\pm2.7$ \\\hline
   & PST & $40.04\pm0.4$ & $24.58\pm1.1$ & $11.4\pm0.5$ & $99.95$ & $53.7\pm2.5$ & $-61.4\pm2.3$ \\\hline

\end{tabular}

\end{table}

Similar analysis was performed for individual subject and it was found that the equation (1) represented quite well with $R^2 \sim0.999$. The values of L and Q for individual subject for pre and post-stimulation were depicted in Fig.4. Except for few (five to six) subjects the slope and magnitude of curvature increased after gyrosonic stimulation. Earlier study on subjects with cardiac illness had shown that the curvature of SD1 and SD12 curves were much reduced compared to those for normal subjects [17]. Based on this result it can be argued that the augmentation of the curvature due to gyrosonic stimulation can then be taken as an indicator of better cardio-dynamics. The gyrosonic may be acting as a 'reward' signal for improvement of sympathovagal balance.\\
The coefficient $\alpha$ and its distribution of detrended fluctuation analysis for subjects was plotted in Fig.5a and 5b respectively for pre and post state of stimulation. The gyrosonic stimulation produced a decrease in the coefficient $\alpha$ for most of subjects and mean for the group is lowered. The mean of $\alpha$ for group changes from     $0.93\pm0.02$ to $0.82\pm0.02$  after stimulation with a p value of $0.003$.
It is to be noted that no significant changes were found in the above mentioned parameters in a group of subjects when we used various monotonous single frequency sound stimulation instead of gyrosonic stimuli.

\section{Conclusions}
The ECG data for short duration ($10$ min) is used to analyze the heart rate variability in time and frequency domain in order to assess the influence of novel gyro-sonic stimulation. It is found that the values of parameters SD1, SD2 and their ratio $SD12$ that quantify the Poincare$\acute{}$ analysis of $RR_N$ interval become higher in post stimulation. The variation of these parameters with lag number is represented by an equation that fits excellently the data of SD's for group average and individual. The coefficients of linear and quadratic (curvature) term of $SD12$ (and $SD1$) vs M relationship are enhanced due to gyrosonic stimulation.The equation (1)  is an important findind as it provides the quantitative maesure of variation of SD12 with M - in particular the curvature of the plot which as appears from this and earlier [18] can be taken as a good measure of state of cardio-dynamics.The coefficient $\alpha$ of DFA  is also decreased in post stimulation state. Gyrosonic stimulation of short duration reduces the heart rate and enhances SD and RMSSD.  In frequency domain both LF and HF power are higher in post stimulation condition.   The trend of the changes of most of the indicators of HRV supports the hypothesis that gyrosonic has the capacity to influence sympatho-vagal regulation in a favorable way and the gyrosonic can be considered as reward signal for improvement of autonomic activity. The importance of sympatho-vagal regulation on various diseases has already established.So, there is a possibility that the gyrosonic may act through improvement in autonomic regulation. However, more work on a larger group of patients with different health conditions are needed for establishing all aspects of gyrosonic.

\section{Acknowledgement}

The authors gratefully acknowledge the Biomedical Signal analysis group, Department of Applied Physics ,Univ. of Kuopio, Finland for allowing to use HRV software, Dr S. Mazumdar for his critical comments and Mr S. Ghosh for his technical help.

%\begin{thebibliography}{letter}

\section{References}

\noindent [1] Recordati G. A thermodynamic model of the sympathetic and parasympathetic nervous systems.  Autonomic Neuroscience: Basic and Clinical 2003; 103: 1-12.

\noindent [2] Chambers AS, Allen JJB. Cardiac vagal control, emotion, psychopathology, and health . Biological Psychology 2007; 74: 113-115.

\noindent [3] Curtis BM,O'Keefe JH. Autonomic Tone as a Cardiovascular Risk Factor: The Dangers of Chronic Fight or Flight. Mayo Clinic Proceedings 2002; 77: 45-54.

\noindent [4] Iwananga M, Moroki Y. Subjective and physiological responses to music stimuli controlled over activity and preference.  Journal of Music Therapy1999;   36: 26-38.

\noindent [5] Blair RW and Thompson GM. Convergence of multiple sensory inputs onto neurons in the dorsolateral medulla in cats.  Neuroscience 2002; 67 : 721-729.

\noindent [6] Salamon E, Kim M, Beaulieu J, Stefano GB. Sound therapy induces relaxation : down regulating stress processes and pathologies. Med Sci Monit 2002; 9 : RA96-RA101.

\noindent [7] Griffiths TD, Bench CJ and Frackowiak RS.  Human cortical areas selectively activated by apparent sound movement.  Curr. Biol 1994; 4: 892-895.

\noindent [8] Xiang J, Chuang S, Wilson D, Otsubo H, Pang E, Holowka S, Sharma S, Ochi A, Chitoku S. Sound motion evoked magnetic fields Clinical Neurophysiology 2002;113:  1-9.

\noindent [9] Griffiths TD, Rees G, Rees A, Green GG, Witton C, Rowe D, Buchel C, Turner R, Frackowiak RSJ. Right parietal cortex is involved in the perception of sound movement in humans.  Nature Neuroscience 1998; 1: 74-79.

\noindent [10] Bandopadhyay S, Mandal MK, Chakrabarti PP, Ghatak SK, Chowdhury R, Ray S. Moving sound reduces arousal in psychosomatic patients.   Intern. J. Noeuroscience 2006 ; 116 : 915-20.

\noindent [11] Malliani A, Pagani M, Lombardi F, Cerutti S. Cardiovascular neural regulation explored in the frequency domain.Circulation 1991; 84:  482 - 492.

\noindent [12] Hayano J, Skakibara Y, Yamada A, Yamada M, Mukai S, Fujinami T, Yokoyama K, Watanabe Y, Takata K. Accuracy of assessment of cardiac vagal tone by heart rate variability in normal subjects. The American Journal of
Cardiology 1991; 67: 199-204.

\noindent [13] Task Force of the European Society of Cardiology and the North American Society of Pacing and Electrophysiology. Heart Rate Variability: standards of measurement, physiological interpretation, and clinical use. Circulation 1996; 93(5): 1043-65.

\noindent [14]Woo MA, Stevenson WG, Moser DK, Trelease RB, Harper RM. Patterns of  beat -to-beat heart rate variability in advanced heart failure.  Am Heart J1992; 123: 704-10.

\noindent [15] Tulppo MP, Makikallio TH, Takala TE, Seppanen T, Hurikuri HV. Quantitative beat-to-beat analysis of heart rate dynamics during exercise.  Am J Physiol Heart Circ Physiol 1996; 271: H244-H252.

\noindent [16] Brennan M,  Palaniswami M, Kamen P. Do existing measures of Poincare plot geometry reflect nonlinear features of heart rate variability?. IEEE Trans. Biomed. Eng. 2001; 48: 1342-1347.

\noindent [17] Lerma C, Infante O, Perez-Grovas H, Jose MV. Poincare plot indexes of heart rate variability capture dynamic adaptations after haemodialysis in chronic renal failure patients. Clin Physiol. Funct. Im 2003; 23: 72-80.

\noindent [18] Thakre TP, Smith ML. Loss of lag-response curvilinearity of indices of heart rate variability in congestive heart failure. BMC Cardiovascular Disorders 2003; 6:  7.

\noindent [19] Contrearas P, Canetti R, Migliaro ER. Correlation between frequency-domain HRV indices and lagged Poincare plot width in healthy and diabatic subjects. Physiol  Meas 2007; 28: 85.

\noindent [20] Shi P, Zhu Y, Allen J, Hu S. Analysis of pulse rate variability derived from photoplethysmography with the combination of lagged Poincare plots and spectral characteristics. Medical Engineering and Physics 2009; 31:
866-871.

\noindent [21] Bak P, Tang C, Wiesenfeld K. Self-organized criticality: An explanation of the 1/f noise.  Phys. Rev. Lett 1987; 59: 381-384.

\noindent [22] Peng CK, Havlin S, Stanley HE, Goldberger AL. Quantification of scaling exponents and crossover phenomena in nonstationary heartbeat time series. Chaos 1995; 5:  82-87.

\noindent [23] Rodrigiuez E, Lerma C, Echeverria JC, Alvarez-Ramlrez J. ECG scaling properties of cardiac arrhythmias using detrend fluctuation analysis. Physiol. Meas 2008; 29: 1255-1266.

% \end{thebibliography}

\newpage
\begin{figure}[htb]
\begin{center}
    \epsfig{file=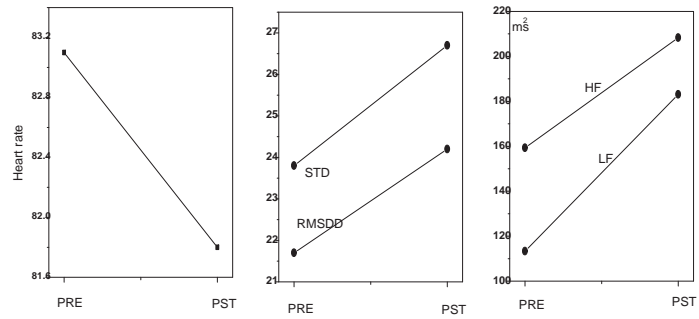,width=10cm}
    \caption{Change of mean heart rate (HR),mean SD,mean RMSSD in RR-interval and mean Power of LF and HF  before (PRE) and after (PST) gyrosonic stimulation.                               For all parameters the change is significant at p$<0.05$.}

\end{center}
\end{figure}

\newpage
\begin{figure}[htb]
\begin{center}
    \epsfig{file=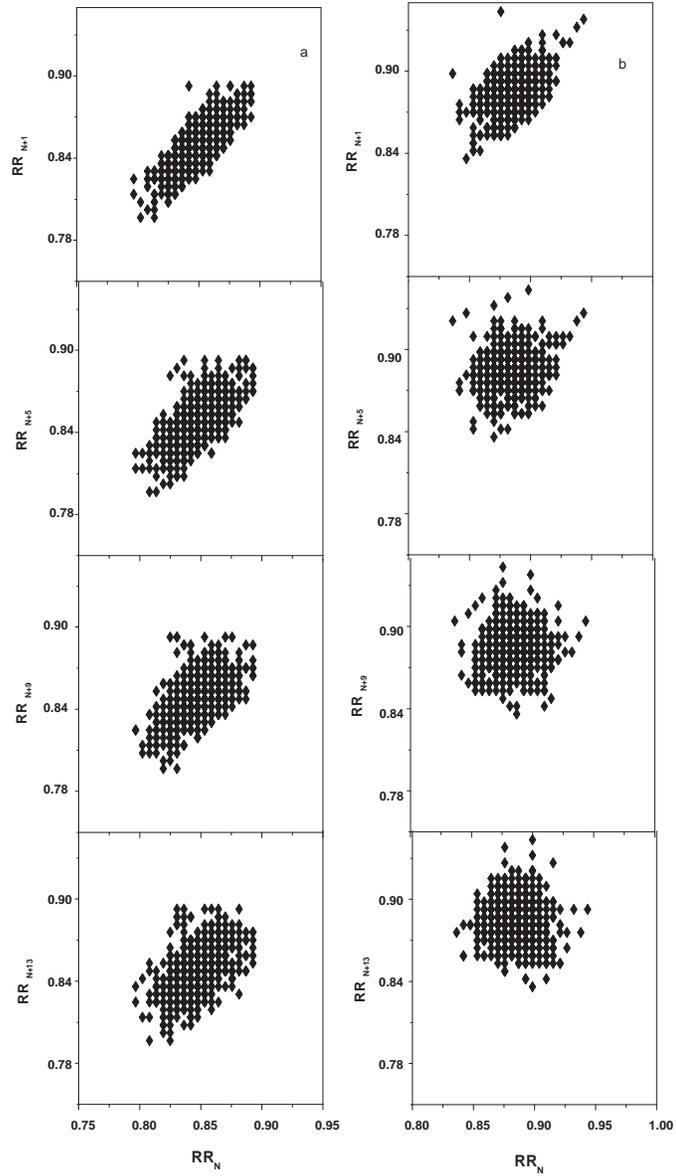,width=10cm}
    \caption{Poincare plot $RR_{N+M}$ vs $RR_N$ of one subject before (left) and after (right) gyrosonic stimulation for $M=1,5,9 $ and $13$.}

\end{center}
\end{figure}

\newpage

\begin{figure}[htb]
\begin{center}
    \epsfig{file=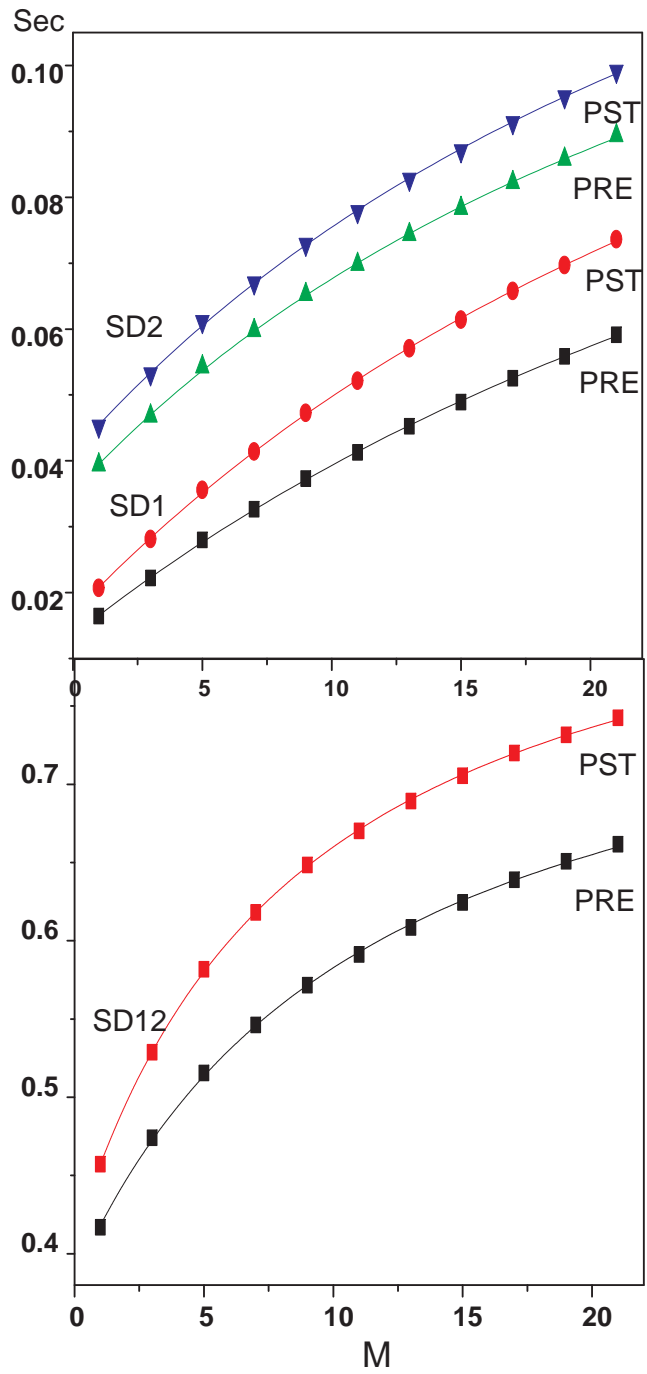,width=10cm}
    \caption{Variation of  mean SD1 and mean SD2 (upper) and mean SD12 (lower) with lag number M before (PRE) and after (PST) gyrosonic stimulation.}

\end{center}
\end{figure}

\newpage
\begin{figure}[htb]
\begin{center}
    \epsfig{file=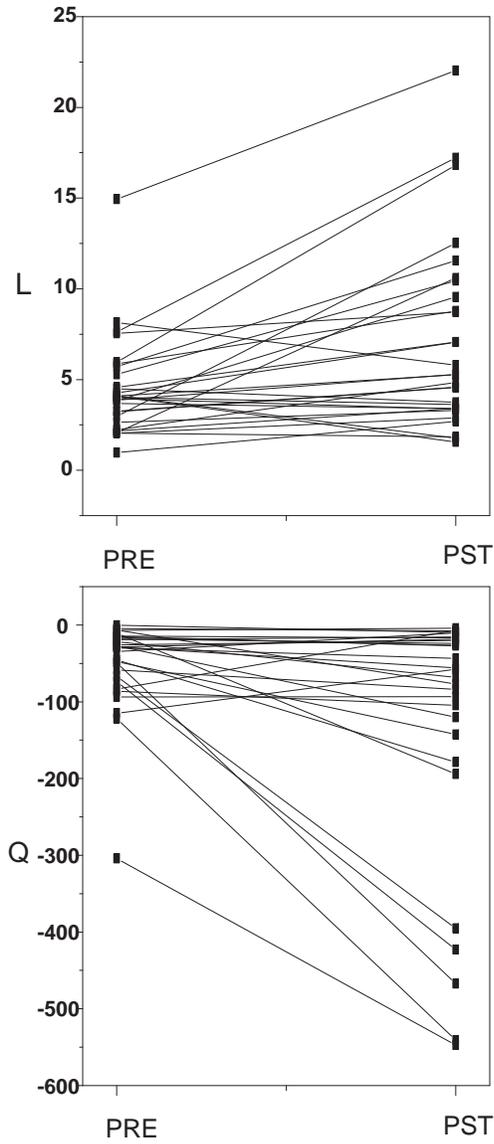,width=8cm}
    \caption{Plot of slope L and curvature Q for individual subject before (PRE) and after (PST) gyrosonic stimulation. Unit of L and Q is in $10^{-2}$s.}
\end{center}
\end{figure}

\newpage
\begin{figure}[htb]
\begin{center}
    \epsfig{file=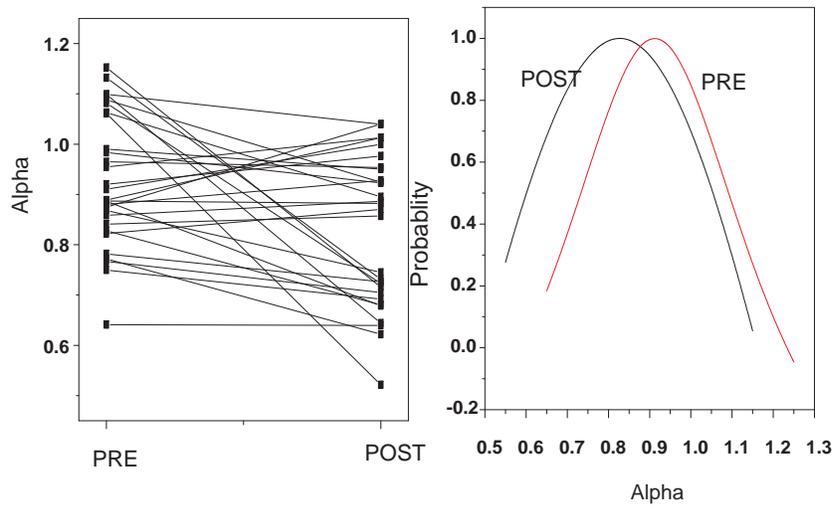,width=12cm}
    \caption{The DFA coefficient $\alpha$  of individual subject (left) and the probability distribution of $\alpha$ (right fig.) before (PRE) and after (PST) gyrosonic stimulation.}
\end{center}
\end{figure}

\end{document}